\documentclass[runningheads,a4paper]{llncs}
\usepackage{amssymb}
\usepackage{graphicx}
\usepackage{caption}
\usepackage{subfig}
\usepackage{float}
\usepackage{wrapfig}
\usepackage{listings}
\usepackage{paralist}
\usepackage{amssymb}
\usepackage{amsmath}
\usepackage{url}
\usepackage{color}
\usepackage{hyperref}
\newcommand{\keywords}[1]{\par\addvspace\baselineskip\noindent\keywordname\enspace\ignorespaces#1}

\begin{document}
\mainmatter
\title{The Hidden Internet of Iran}
\subtitle{Private Address Allocations on a National Network}
\author{Collin Anderson\thanks{This project received partial funding from the Center for Global Communication Studies at the University of Pennsylvania's Annenberg School for Communcation.}}
\titlerunning{The Hidden Internet of Iran}
\authorrunning{Collin Anderson}
\institute{
  \texttt{\href{mailto:collin@averysmallbird.com}{collin@averysmallbird.com}}
}
\maketitle
\begin{abstract} 
While funding agencies have provided substantial support for the developers and vendors of services that facilitate the unfettered flow of information through the Internet, little consolidated knowledge exists on the basic communications network infrastructure of the Islamic Republic of Iran. In the absence open access and public data, rumors and fear have reigned supreme. During provisional research on the country's censorship regime, we found initial indicators that telecommunications entities in Iran allowed private addresses to route domestically, whether intentionally or unintentionally, creating a hidden network only reachable within the country. Moreover, records such as DNS entries lend evidence of a `dual stack' approach, wherein servers are assigned a domestic IP addresses, in addition to a global one. Despite the clear political implications of the claim we put forward, particularly in light of rampant speculation regarding the mandate of Article 46 of the `Fifth Five Year Development Plan' to establish a ``national information network,'' we refrain from hypothesizing the purpose of this structure. In order to solicit critical feedback for future research, we outline our initial findings and attempt to demonstrate that the matter under contention is a nation-wide phenomenom that warrants broader attention.
\keywords{censorship, national internet, Iran, rfc1918}
\end{abstract}
\section{Introduction}
\label{sec:introduction}

The primary purpose of this paper is to document that Iran has broken from commonplace Internet addressing standards to create a private network that is only accessible within the country. We attempt to establish that this private network is  accessible to a wide section of the nation's Internet users and then begin to outline our initial findings in order to provoke broader discussion and solicit feedback on our claim. This publication should be considered a preface to a broader study on the nature of Iran's information communications infrastucture and censorship regime and, as such, where possible we will defer larger questions on such topics until a future occassion. Furthermore, to the extent possible, we focus our assessment on that which is qualitatively measurable, and limit attempts to augur the political aspects of the matters at hand. This paper is not intended to be comphrensive, and we err on the site of brevity where possible. 

Toward these ends, our contributions are threefold:

\begin{enumerate}
    \item We establish that there is a coordinated decision by a subset of Iranian Internet Service Providers (ISP) and government agencies to adopt the use of private Internet Protocol (IP) addresses across networks.
    \item We begin to identify the participants in this arrangement and map logical neighborhoods within the private space.
    \item We attempt to enumerate the services and resources that are available on this hidden network.
\end{enumerate}

The majority of the experiments outlined are motivated toward collecting initial, open-ended data on an unexpected phenomenon; where possible our results are publicly available for outside investigation at: \url{http://github.com/collina/Filternet}

The remainder of this paper is structured as follows. In Section \ref{sec:setup}, we describe a `kitchen sink' approach used to extract information from a variety of sources on the nature of the network. Fundamental principles of network addressing standards are briefly discussed in Section \ref{sec:rfc1918}; they are then applied to the case of Iran in Section \ref{sec:analysis}. Section \ref{sec:depth} attempts to demonstrate that the reachability of networks is not a localized phenomenom and presents evidence that these qualities are the product of intentional design. Data collection methods used to described the content on the network are discussed in Section \ref{sec:content}. Finally, the paper is concluded in Section \ref{sec:conclusion} with an enumeration of unresolved questions.

\subsection{Note About Rhetoric}\label{sec:rhetoric}

The claims of this paper are contemporanous to widespread fears regarding the future of Iran's Internet, particularly whether the government will shift from a strategy of aggressive filtering to blocking all access to foreign sites. We can factually state that the `Fifth Five Year Development Plan of the Islamic Republic of Iran (2010-2015)'\cite{article46} mandates the development of a `national information network' for the purposes ``e-government services, industry, information technology, information literacy, and increased productivity in the areas of economic, social and cultural activities.'' The development plan continues on to describe the need for secure and private communications between government ministries, businesses and the public, based on increased access to broadband connectivity and investments in national data centers.

After the codification of the development plan, Reza Taghipour, head of the Ministry of Information and Communications Technology, and other government officals began a public campaign extoling the virtues of a national Internet, promising domestic alternatives for email and search, while warning of the dangers of foreign services. 

\begin{quote}
``a genuinely halal network, aimed at Muslims on an ethical and moral level'' - Ali Agha-Mohammadi, Deputy Vice President for Economic Affairs \cite{wsj:AghaMohammadi}
\end{quote}
\begin{quote}
``A national internet can be very effective to protect the country's information and the people's security.'' - Esmail Ahmadi Moghaddam, Chief of Law Enforcement Forces \cite{fars:Moghaddam}
\end{quote}

While a substantial portion of the narrative in the Western press and Persian blogs was based on worst-case scenarios and hoaxes, the parallels between official comments and the presence of a private, domestic network are stark. However, upon subsequent investigation, we found reports of the national use of private addresses as far back as January 2010. At that time the Statistical Center of Iran had requested the public to report information such as income and property value via the site `amar.org.ir.' It was reported\cite{khabar,technet} then that the domain resolved to the IP address 10.10.33.40 and that the site self-reported 2 million visits.\footnote{At the time of writing, a DNS query to Google's Public DNS returns the address 217.218.11.61 for amar.org.ir and not host appears to be attached to 10.10.33.40.} Additionally, the most visited Iranian site, the filtering page returned within the country at 10.10.34.34, was mentioned as early as June 24 2010.\footnote{It is noteworthy that at least one blog stated in September 2011 that a number of unnamed VPN clients were found to connect to 10/8 addresses. While the claim put forward is not subtantiated by evidence, the ownership and end destination of popular VPN networks is a matter worth investigation.} Therefore, we assert neither that the matters described herein are new, nor that they indicate immediate plans to disconnect from the global Internet.

\section{Standards on Private Internet Addressing}
\label{sec:rfc1918}

The fundamental basis of network communications is the unique assignment of IP addresses grouped together within logical subdivisions, refered to as subnets, to hosts within the network. Globally, the uniqueness property of network numbers, such IP addresses or Autonomous Systems numbers (ASN), necessitates a central authority for allocation and coordination to prevent conflicts. These responsibilities are delegated to a private governing body, the Internet Assigned Numbers Authority (IANA) and its regional representatives, from which ISPs and other entities obtain blocks of addresses. 

However, since the potential address space of a 32-bit IP scheme is finite and not every host requires direct bidirectional accessibility on the Internet, IANA has reserved three blocks of IP addresses that may be used for private, local networks without requiring coordination or approval. In the standards document outlining the use of these private address\cite{rfc1918}, Rekhter, et al differentiates private and public hosts thusly:

\begin{itemize}
    \item \emph{Private Hosts} Hosts that do not require direct reachability within the global Internet; or only require a limited set of outside services, which can be handled by intermediary gateways. Private hosts then may maintain IP addresses that are unambiguous within networks, but ambiguous to the broader Internet.
    \item \emph{Public Hosts} Hosts that require access outside the enterprise and maintain an IP addresses that is globally unambiguous and logically reachable.\cite[p.g. 3,4]{rfc1918}
\end{itemize}

Figure \ref{lst:rfc1918blocks} lists the established set of deregulated, reusable blocks of IPs, allocated to help conserve the finite pool of global addresses and make it easier to establish smaller, localized networks of hosts. 

\begin{figure}
\begin{lstlisting}[basicstyle=\footnotesize\ttfamily]
     10.0.0.0     -  10.255.255.255  (16,777,216 Addresses)
     172.16.0.0   -  172.31.255.255  (1,048,576 Addresses)
     192.168.0.0  -  192.168.255.255 (65,536 Addresses)
\end{lstlisting}
\caption{IANA Reserved Private Address Blocks}
\label{lst:rfc1918blocks}
\end{figure}

Those hosts assigned IP addresses within a private address block may communicate within the same network, however, they may not directly exchange traffic with the broader Internet. RFC1918 makes explicitly clear the expectations vis-\`a-vis routing traffic from hosts with private addresses.

\begin{quote}
   Because private addresses have no global meaning, routing information
   about private networks shall not be propagated on inter-enterprise
   links, and packets with private source or destination addresses
   should not be forwarded across such links. Routers in networks not
   using private address space, especially those of Internet service
   providers, are expected to be configured to reject (filter out)
   routing information about private networks. If such a router receives
   such information the rejection shall not be treated as a routing
   protocol error.\cite[p.g. 5]{rfc1918}
\end{quote}

Traditional networks utilize the mechanism of `network address translation' (NAT) to allow globally-accessible intermediaries to act as gateways, transparently sending and receiving traffic to the broader Internet on behalf of private hosts within the same network. As the number of directly connected Internet devices has increased, the pool of unallocated IP addresses available for networks has diminished; many ISPs have employed techniques such as `Carrier-Grade NAT' (CGN) or `Large-Scale NAT' in order to stretch allotments. Similar to NAT within an enterprise, in a CGN network, a localized set of users share a single public address, while each user is assigned a private address internal to the network. These addressing schemes have also been leveraged to limit the availability of exclusive content, such as Internet Protocol Television (IPTV), to consumers and decrease the exposure of infrastructure to outside actors.

\section{Experimental Setup}\label{sec:setup}

Before describing our findings, it is necessary to address various ethical aspects of our research and then outline the methods with which we were able to conduct our analysis. Lastly, we close this section by noting some of the shortcomings of this approach and addressing how we attempted to mitigate potential problems.

\subsection{Legal and Ethical Aspects}

During the process of preparing and running our experiments, we took special care to not knowingly violate any laws or, considering the diminishing opportunities for international collaboration or free expression, expose individuals within Iran to potential harm. In addition, all our experiments were in accordance with any pertinent terms of service and within reasonable considerations of network usage, taking care to not engage in behavior that would be considered intrusive. Within a literature review of computer science research, we found that widescale surveying of accessible networks has become an accepted manner of collecting data on aggregate qualities of the Internet or communications infrastructure\cite{effssl,wustrow,clayton2006ignoring}. In line with other works, our data collection is limited to the normal, expected functions of remote system that are reachable without requiring credentials. 

\subsection{Points of Observation}\label{sec:vantage}

To ensure that our findings are not a localized phenomenom, we have sought to obtain a heterogenous set of vantage points on different logical segments of Iran's communications infrastructure for the purposes of measurement and observation. 

\textbf{Host 1, Tehran}: The bulk of our initial experimentation was conducted from a host in Tehran, Iran that falls under a significant provider, who services state agencies such as the Islamic Republic of Iran Broadcasting. While outside the scope of this paper, Host 1's primary upstream peer is Russian Rostelecom (AS12389) through the Information Technology Company (AS12880).

\textbf{Host 2, Tehran}: Secondary testing was performed from a network that provides upstream access for several universities, the Ministry of Commerce and research institutions affiliated with the Ministry of Science, Research, and Technology. The primary upstream provider for Host 2 is Delta Telecom (AS29049) through the Institute for Research in Fundamental Sciences (AS6736).

\textbf{Open HTTP Proxies}: In order to test as wide of a subset of Iran's networks as possible, we enumerated a pool of openly accessible HTTP proxies, on both public and private addresses, and made requests to a mixed set of public and private hosts through them. For our limited purposes, we assumed that if the proxy was able to successfully relay requests, some level of connectivity existed between the intermediary and the destination. After scanning publicly-accessible Iranian IP blocks for servers listening on the port used by the Squid web cache service, around a hundred proxies were identified on 27 networks.\footnote{We use ASN as a stand-in for grouping networks, recognizing that this creates incongruenties in reporting, since some resources report these details with differing granularity.} The same process repeated on private address space yielded 15 proxies on indeterminate networks.

\subsection{Shortcomings}

There are natural limitations imposed on research from a small subset of the hosts with a country, particularly where limited outside information exists. Furthermore, we should recognize that we have no reliable information about the owners of our points of observations, such as whether they conduct localized network filtering that differs from other locations with regard to the type, technologies or sophistication. Active analysis of a censorship systems and networks can easily attract attention if no special care is taken. Considering the history of attacks against the public's ability to safely access the global Internet, the ability to coordinate real time sources of data is a vital resource that we have sought to be protect for future use.

We acknowledge the aforementioned shortcomings of our experimental methods, and have stratified our sourcing while attempting to be conservative in our claims.

\section{Analysis}
\label{sec:analysis}

\begin{wrapfigure}{R}[0pt]{0.5\textwidth}
    \vspace{-20pt}
    \parbox{0.5\textwidth}{
      \scriptsize
      \ttfamily
        traceroute to facebook.com (69.63.181.12)
        \ldots Home Network\ldots\\
        2 91.99.***.***.parsonline.net [91.99.***.***]\\
        3 \textcolor{red}{10.220.1.2}\\
        4 2.180.2.1\\
        5 217.219.64.115\\
        6 78.38.245.6\\
        7 78.38.245.5\\
        8 78.38.244.242\\
        9 78.38.244.241\\
        10 \textcolor{red}{10.10.53.61}\\
        \ldots Traffic Exits Country\ldots\\
    \label{lst:publictrace}
    \vspace{-5pt}
    \caption{Traceroute to Public Internet}
    }
    \parbox{0.5\textwidth}{
      \scriptsize
      \ttfamily
        traceroute to 10.10.34.34 (10.10.34.34)\\
        \ldots Home Network\ldots\\
        2  81.12.48.89 (81.12.48.89)\\
        3  10.9.27.1 (10.9.27.1)\\
        4  10.30.153.253 (10.30.153.253)\\
        5  217.218.190.26 (217.218.190.26)\\    
        6  78.38.119.30 (78.38.119.30)\\
        7  78.38.119.210 (78.38.119.210)\\
        8  195.146.33.29 (195.146.33.29)\\
        9  10.10.34.34 (10.10.34.34)\\
    \label{lst:traceroute_filter_page}
    \vspace{-5pt}
    \caption{Traceroute to Content Filter}
    }
\end{wrapfigure}
The most immediate indicators of the national use of private addresses in Iran lie within the filtering regime and international routing paths. Iran maintains one of the most aggressive filtering regimes in the world, blocking access to a wide range of content, determined to offend the political order, religious morality or social norms. Attempts to visit restricted content results in redirection to a site offering suggestions for state-sanctioned content. While the user appears to remain on the site they attempted to visit, the redirection is made within a frame to the private IP address 10.10.34.34 (See Figure \ref{fig:filterresponse}, the results of a request for Facebook.com). \footnote{We defer elaboration on the mechanism until our broader discussion of Iran's filtering regime, however is it worth noting that there appears to be two observable differences in responses, based on the filter rule triggered (`Invalid Site'/`Invalid Pattern') and based on location (`F1-IPM'/`M3-5')}

This censorship mechanism is in part a product of the centralization of Iran's network around the Telecommunication Company of Iran (TCI). All internationally-bound traffic is routed through either a subsidary of the TCI, the Information Technology Company of Iran (AS12880), or the Institute for Studies in Theoretical Physics and Mathematics (AS6736). For the majority of consumer networks, which connect through TCI, the final hop before traffic exits the country appears to be handled by one of at least three Huawei Quidway NetEngine80E core routers that have private network addresses within the 10.10.53.0/24 range (See Figure \ref{lst:publictrace}).

\begin{figure}
\begin{lstlisting}[basicstyle=\scriptsize\ttfamily]
<html><head><meta http-equiv="Content-Type" content="text/html; charset=
windows-1256"><title>F1-IPM</title></head><body><iframe src=http://10.10.34.34?
type=Invalid Site&policy=MainPolicy " style="width: 100%; height: 100%"
scrolling="no" marginwidth="0" marginheight="0" frameborder="0" vspace
="0" hspace="0"></iframe></body></html>
\end{lstlisting}
    \caption{Response to Filtered GET Request}
    \label{fig:filterresponse}
\end{figure}

\begin{wrapfigure}[15]{R}{0.5\textwidth}
  \footnotesize
  \begin{tabular}{|l|l|}
    \hline
      Service (Port) & Number of Host \\
    \hline
      FTP (21) & 12672\\
      SSH (22)  & 8029\\
      Telnet (23)  & 20060\\
      SMTP (25)  & 183\\
      DNS (53)  & 2510\\
      POP (110)  & 78\\
      HTTP (80)  & 9960\\
      IMAP (143)  & 44\\
      HTTPS (443)  & 1366\\
      HTTP-Alt (8080)  & 601\\
    \hline
    \end{tabular}
  \caption{Services on Private IP Space}
  \label{table:hosts_serices}
\end{wrapfigure}

While it is within standard expectations for a network infrastructure to use private addresses internally for reasons of security, flexibility and resource limits, under RFC1918, hosts should not be reachable from outside the immediate network by these private addresses. In order to measure the extent that this private network space is used by the TCI and others, we attempted to make connections to all 16,777,216 possible addresses within the IP block 10.0.0.0/8. 

We performed our data collection in three phases: discovering IP addresses accepting connections on TCP port 23 (Telnet), 53 (DNS) or 80 (HTTP); performing a simple handshake and storing the presented response; and mapping routers to determine logical grouping and routes. Based on the pool of potential networks we narrowed this range to 45928 potential hosts.

\subsection{Measuring Reachability to the Private Network}\label{sec:depth}

As acknowleged in Section \ref{sec:setup}, we are naturally limited in the extent to which we can assert that our findings are valid for the entirity of the country's domestically routable networks. Even according the public statements made by the Ministry of Information and Communications Technology, the implementation of the mandates of Article 46 is incomplete and is being progressively deployed, prioritizing academic institutions and government ministries. Therefore, we do not expect access to be universal or consistent across all geographic regions or networks. 

Using our established pool of open, globally-accessible proxies, we attempted to make proxied HTTP GET requests to: \begin{inparaenum}[(i)]
    \item a globally available website located inside of the country, 
    \item a domain pointing to a private IP address, and 
    \item a private IP address running a webservice,
    \item an a unfiltered global website located outside of the country, as a control to ensure the proxy is properly working. 
\end{inparaenum}\footnote{In order: peyvandha.ir, ou.imamreza.ac.ir, 10.8.12.18 (what we suspect is the private address of the national email service), google.com (which during testing was unfiltered)}
If any of these requests returned a ``200 OK,'' the standard response for successful HTTP requests, and matched our the expected page title, then the end destination was counted as reachable. In parallel, this experiment was repeated with our set of proxies with private addresses. Using this process, we could verify that 27 distinct networks, and 12 hosts with private addresses, were able to reach at least one of our private space hosts. Relying on proxy servers to measure connectivity to private networks is subject to a number of hypothetical scenarios under which misconfiguration on the part of the administrator, or limitations on intended use, may trigger a false negative. A false positive is possible, however, the foreseeable circumstances under which this may occur, since we match the integrity of the response data against expected results, seem less probable. Thus, we focus on successful connections as a measure of connectivity and do not give failures  much weight.

The results of this measurement are enumerated in Listing \ref{lst:reachability_proxy}. Of the publicly accessible networks, 24 (89\%) had at least one host that was able to connect to Iran.ir's private IP address and 21 (77\%) were able to connect to the Imam Reza International University. For internal proxies, 13 were able to connect to Iran.ir and 6 could reach the university.\footnote{Successfully reaching `ou.imamreza.ac.ir' is a more difficult task as it required the proxy to successful resolve a domain against an available nameserver.}

\section{How Widely Used is the Private Network}\label{sec:content}

Unlike the global Internet\cite{rir_allocations}, no public registry of the networks attached to this private address space exists. Therefore, we sought to leverage a mixed methods approach of mining open sources to begin a very preliminary mapping of the hidden Internet. Through data collection methods, such as collecting service banners and tracing the path of traffic, we were able to learn the public identities of hosts within the private address, describing the:
\begin{inparaenum}[(i)]
    \item logical neighborhoods of networks,
    \item purposeful utilization of private networking spaces, and 
    \item wide coordination of network infrastructure.
\end{inparaenum}
A significant number of these responses included a fully qualified domain name (FQDN) and public addresses that could be matched against public records. Using this methodology, we begin to build a rudimentary understanding of the logical ownership and participation in the national network, which is compiled together in Figures \ref{table:nation_network_sites} and \ref{lst:private_address_domains}.

\begin{wrapfigure}[6]{R}{0.5\textwidth}
    \scriptsize
    \ttfamily
    \begin{minipage}{0.5\textwidth}
        \$ nc 10.143.177.18 25\\
          220 webmail.isfidc.com ESMTP MailEnable\ldots\\
        \$ dig +short webmail.isfidc.com\\
          91.222.196.18\\
    \label{lst:banner_grabbing}
    \caption{Banner Grabbing on SMTP}
    \end{minipage}
\end{wrapfigure}

\subsection{Service Banners}\label{sec:banners}

During the enumeration of hosts outlined under Section \ref{sec:analysis}, we sought to identify Internet information services responding within private network spaces, particularly web, email and domain servers. After identifying which of a possible 16,777,216 IP addresses were populated with hosts, we then were able to conduct the secondary retrieval of data and attempt to assess the amount of web content with the private address space.  

Considering the scale of such a search, determination that a particular service was available was based on a simple acknowledgement by the remote host that some application was listening on the port commonly associated with it.\footnote{We recorded a host as interesting based on it responding to a TCP SYN request.} Depending on the type of web service, a number of variables could skew the degree to which our results match the effective reality. For example, many home DSL routers will allow device configuration through a web or telnet frontend, which would be difficult to differentiate from a complete Internet site. Moreover, companies commonly host multiple websites on a single server; a simple connection would not divulge the amount of content available.\footnote{The occurrence of server names such as `Sepehr Server Farm' on the network 10.30.74.0/24 lend evidence of shared hosting servers on private space.} Bearing in mind these caveats, Figure \ref{table:hosts_serices} details the results of this search. 

As noted, while ownership of IP segments is public record, no such resource is available for unregulated, private addresses. Whether or not such a scan is an accurate representation of the richness of this environment, it does facilitate the exploration of the ownership of its logical subdivisions by linking the content hosted on networks with particular organizations or public network blocks. As demonstrated in Figure \ref{lst:banner_grabbing}), many services divulge what they believe to be their identities by simply connecting to them. Information such as a FQDN can be linked to a publicly available IP address; this link could then be validated by matching secondary information such as timestamps and software versions. After ensuring that 10.143.177.18 and `91.222.196.18' are most likely the same server, we can reasonably assume that some subnet that contains the address 10.143.177.18 belongs to the `Computer Research Center of Islamic Seminary of Isfahan.'\footnote{The Research Center is assigned the IP Block 91.222.196.0/22 attached to AS48121.} Even where information retrieved is incomplete, context regarding location or content narrows the possibilities for further investigation, such as matching the content of the responses of private webservers with sites that have public components crawled by a search engine. We might also find indicators in determining where the majority of links within returned sites lead. 

\subsection{DNS Records}

A particularly striking aspect of our findings is the occurrence of domains that maintain DNS records for valid, reachable private IP addresses, or multiple records that include both public and private pointers. Evaluation of all possible DNS records is difficult since IRNIC, the maintainer of the Iranian country top level domain \emph{.ir}, does not appear to release the zone file that would list registered domains. We are therefore limited to examining public resources and service banners to build Figure \ref{lst:private_address_domains}.

\subsection{NAT Traversal and ICMP Deflection}\label{sec:deflection}

To communcate bidirectionally access the Internet machines assigned private address will either have to have an additional, public address or be able to transit through a NAT gateway. Since we are interested greatly in international accessibility, as well as correlating private with public addresses, but unable to directly query private space hosts, we attempt to make an ICMP Echo request from within the country to a private address, forging the source address as our server outside of the country. The destination should issue an ICMP Echo-Reply to the external host. If the response travels through a gateway along its route, the private address of the remote host will be rewritten with the public address of the intermediary, lending evidence to network ownership.

ICMP Echo requests were sent to all remote hosts previously identified, using the sequence data field to correlate the private address queried with the response seen. There are a number of reasons why a host would not reply, including firewall policies or changes in reachability since the scan. Out of the 45,928 hosts queried, our external observer recieved 10,344 responses, of which 358 had a public source address. Another 408 replied with a different private address than that queried, which may be due to reasons such as the response being from an intermediary along the route or the remote host having multiple addresses attached. The remainder of replies maintained the private address that was queried. The vast majority of public addresses seen resolved to the networks of TCI's Information Technology Company, with the remaining from AsiaTech Inc. (2 hosts), Soroush Rasanheh Company Ltd (8 Hosts), CallWithMe (1), Neda Rayaneh (6). \footnote{There were two abnormalities, likely the result of network addressing errors by administrators, where public addresses appeared to originate from France Telecom and AT\&T Services, Inc.}

\subsection{Traceroute Data}

Tracing the route of network traffic is trival and lends evidences as to the participation in the private addressing scheme. Using the example of the filtered site page (10.10.34.34), Figure \ref{lst:traceroute_filter_page} shows the route that a request takes to arrive at its destination. We note that the IP address immediately preceding the final destination (195.146.33.29) is publicly registered to `Data Communication Affairs,' a subdivision of the Information Technology Company. Since network traffic will route to the datacenter in which the host is located before reaching the end destination, we may infere that Data Communication Affairs has a role in the maintenance of the filtering apparatus. 

We can apply this same theory to other discrete destinations of the network, by tracing the routes taken to every possible IP address beginning with 10. and ending in .1, under the assumption that they represent the smallest reasonable division of the private address space. Using our in-country hosts, we are able to create the path maps in Figures \ref{fig:10dotroutes_khayyam} and \ref{fig:10dotroutes_ferdowsi}. We extend this work by extracting the unique public addresses found in routing and determining their ownership in Figure \ref{table:public_traceroute_addresses}. \footnote{Again, we have removed four instances where networks associated with AOL Transit Data Network (1668) and AT\&T Services, Inc. (AS7018) were found likely due to administrators miscalculating the private 172.16.0.0/12 subnet.}

{
\renewcommand{\arraystretch}{1.3}%
\begin{figure}[h]
    {\footnotesize
      \centering

        \begin{tabular}{|l|l|}
            \hline
                IP Address & Host/Network \\
            \hline
                10.8.12.18 & Iran.ir National Webmail Service \\
                10.8.218.0/24 & Pishgaman, ADSL Internet Service Provider \\
                10.10.34.34 & Data Communication Affairs's Filtered Site Page \\
                10.10.36.0/24 & Telecommunications Company of Iran \\ 
                10.30.54.0/24 & Parsonline, ADSL Internet Service Provider \\
                10.254.50.0/24 & Islamic Republic of Iran Broadcasting \\
                10.9.28.0/24 & Islamic Republic of Iran Broadcasting \\
                10.143.218.199 & Telecommunications Company of Isfahan \\
                10.56.59.198 & Khorasgan Islamic Azad University, Isfahan \\
                10.7.234.0/24 & Ministry of Agriculture\\
                10.30.170.0/24 & Ministry Of Education\\
                10.21.243.37 & National Internet Development Agency of Iran\\ 
            \hline
        \end{tabular}
        \caption{Sampled Identifiable Networks and Sites on Private IP Space}
        \label{table:nation_network_sites}
    }
\end{figure}
}

\section{Conclusions and Further Questions}
\label{sec:conclusion}

We have sought to shed light and collect data on a previously unexplored and unorthodox aspect of Iran's information communication infrastructure, then offered evidence that the network design described is the product of purposeful design. Through comparative analyses from a diversity of sources, we have lent evidence to the premise that the national Internet is internally consistent and widely reachable. In addition, while we do not attempt to augur the future of the country's international connectivity, we have offered conjecture that Internet of the Islamic Republic has an increasingly autonomous property at its core. In this capacity, the private space networks we have encountered is consistent with the expectation put forward by the Ministry of Information and Communications Technology. However, we have only dredged up more questions than answers; moving forward, we see an immediate need to answer the following questions:

\begin{description}
\item[Are Private Addressing Schemes Related to IPv4 Exhaustion?] \vspace{1pt}
Based on RIPE delegated address data retrieved September 18 2012, networks registered in Iran have approximately a total of 9,555,968 IPv4 addresses allocated. While based on the number of Internet-connected homes, it is difficult to imagine Iran nearing the point of exhaustion.\footnote{Wide differences exist in the estimation of home Internet users, ranging from 43\% (Broadcasting Board of Governors) to 21\% (International Telecommunication Union, Iranian Statistics Center)\cite{upenn}; on September 4 2012, the consumer ADSL and WiMAX ISP `Gostaresh-e-Ertbata-e Mabna' was awarded one of the last large unallocated blocks of IP addresses remaining.} Considering the multitude of consumer devices with data capacity, it remains difficult to assess the overall necessity of CGN-type solutions. However, the scenario outlined herein differs significant from the typical CGN scenario in that routes are shared outside of local networks. Furthermore, CGN is generally employed to mitigate the impact of residential or mobile Internet connectivity, rather than government ministries, universities or content hosts.
\item[Does the Private Network have Global Internet Access?] In Section \ref{sec:deflection}, we attempted to test the public accessibility of the private space hosts by forging ICMP echo requests to hosts outside of the country. Additionally, similar to the setup used in Section \ref{sec:vantage}, we tested the reach of proxies listening on private addresses. In both cases, it did appear that a minority of systems queried were able to communicate through either NAT gateways or additional public address attached to the host.
\item[Will Iran Segment the National Network from International Traffic] 
\item[Using DNS?] As we have demonstrated, Iranian organizations have used the domain system to implement national Internet sites with a degree of opacity to the user. Moreover, in some cases auxiliary records exist to allow websites to failover from public to private networks. The logical next step for telecommunications entities would be to implement DNS tampering or `split-horizon' mechanisms to provide different DNS responses, based on whether the query originates in Iran or internationally. In preliminary testing, we have found little evidence that Iran has attempted to interfere with DNS services; instead, filtering of services such as HTTP appears to be done through transparent proxying and other forms of traffic interception. In the process of investigating content hosted on the private network, we found a number of FQDNs that were unresponsive to requests and did not resolve by DNS to an IP. Later we found that a selection of these domains would only properly resolve against an Iranian DNS, such as the public service of IRNIC, as demonstrated in Figure \ref{lst:dns_prop_failure}. Revisiting this matter within a limited subset of domains revealed `blizz.ir' is not the only instance of this phenomenom, with further examples found in the `isftak.ir' and `geeges.co.ir' domains. Whether intentionally or unintentionally, the use of private addresses for DNS nameservers has the strong potential to interfere with normal Internet functions by impeding the global proliferation of route information.
\item[Is the Private Address Space Growing?] The data captured by the project represents a narrow window of observation in late August and early September 2012; we therefore lack a proper perspective on the extent or rate to which this address space is expanding or contracting. We have noted in Section \ref{sec:rhetoric} that the private address space has been in use since 2010. Scanning such a large address block is a significant endeavour and may attract unnecessary attention; therefore, we propose to continue to monitor the routes to the smallest reasonable class C networks, and reserve wider scans to infrequent occassions.
\item[How Much Content is Located Exclusively on the Private Network?] We have attempted to establish that a wide range of public services are duplicated on the private network, or exist solely for internal users. This has been a case based on time-consuming investigation of hints returned from hosts and in no way represents a thorough evaluation of the state of the national network. An opportunity track similar to monitoring address use is in order and will be pursued.
\item[Are Users Given Private Addresses?] In Figure \ref{lst:publictrace}, it appears that the most immediate next hop for a Parsonline user, which would generally be a DSLAM, is a private space address. However, it is unclear whether the address reported to be associated with the host is the product of CGN, or a directly assigned public address. Testing was not performed from residential connections, so it remains unclear whether any ISPs utilize private addresses for hosts. During analysis of content, we did determine informally that the majority of responses on HTTP services where from ADSL modems, which may indicate that some home users at least have a dual set of private and public addresses.
\end{description}

\section*{Acknowledgments}
\label{sec:acks}

This research would not have been possible if not for the substantial contributions of a number of individuals who I am privileged to even know and deeply regret not being able to acknowledge in name; the chilling effect of self-censorship and intimidation is not limited to the borders of a country. Fortunately, it is possible to recognize Fabio Pietrosanti and Arturo Filast\`{o} for their valuable insight on applied network research methods.

\bibliographystyle{plain}
\bibliography{paper}

\appendix
\section*{Appendix}

    \begin{figure}[h]
        \footnotesize
        \begin{tabular}{|l l p{7cm}|}
            \hline
                lib.atu.ac.ir      &  10.24.96.14    & Allameh Tabatabaie University\\
                www.mdhc.ir        &  10.30.5.163    & Vice Presidency for Management Development and Human Capital\\
                www.iranmardom.ir  &  10.30.5.148    & Vice Presidency for Management Development and Human Capital\\
                erp.msrt.ir        &  10.30.55.29    & Ministry of Science, Research and Technology\\
                ou.imamreza.ac.ir  &  10.56.51.27    & Imam Reza University\\
                www.tehranedu.ir   &  10.30.95.7     & Tehran Education Organization\\
                sanaad.ir          &  10.30.170.142  & Private Individual\\
                ww3.isaco.ir       &  10.21.201.50   & Iran Khodro Spare Parts \& After-sales Services Company\\
                iiees.ac.ir        &  192.168.8.9    & International Institute of Earthquake Engineering and Seismology\\
                                   &  169.254.78.139 & \\
                                   &  194.227.17.14  & \\
                                   &  10.10.3.2      & \\
                tci-khorasan.ir    &  217.219.65.5   & Telecommunication Company of Iran, Khorasan\\
                                   & 10.1.2.0        & \\
                adsl.yazdtelecom.ir & 10.144.0.14    & Telecommunications Company of Iran, Yazd\\
                iranhrc.ir          & 46.36.117.51   & Private Individual\\
                                    & 10.30.74.3     & \\
                acc4.pishgaman.net  & 81.12.49.108   & Pishgaman, ADSL Access Provider\\
                                    & 10.8.218.4     & \\
                lib.uma.ac.ir       & 10.116.2.5     & University of Mohaghegh Ardabili\\
                film.medu.ir        & 10.30.170.110  & Ministry Of Education\\
                www.shirazedc.co.ir & 10.175.28.172  & Shiraz Electric Distribution Company\\
            \hline
            \end{tabular}
        \par %
        \caption{Domains, Corresponding A Records and Ownership For Private Addresses}
        \label{lst:private_address_domains}
    \end{figure}

    \begin{figure}[h]
        \footnotesize
\begin{lstlisting}[basicstyle=\scriptsize\ttfamily,captionpos=b]
ASN     (Hosts)  10.8.12.18  google.com     ou.imamreza.ac.ir   peyvandha.ir
-----------------------------------------------------------------------------
RFC1918 (15)     13          9              6                   12
-----------------------------------------------------------------------------
44285   (2)      0           2              0                   0
31549   (3)      0           0              1                   0
50810   (2)      1           0              1                   0
39501   (1)      1           1              1                   1
48159   (4)      1           3              1                   3
50892   (1)      1           1              1                   1
42163   (2)      1           1              2                   1
51235   (1)      1           1              1                   1
16322   (5)      5           3              3                   3
25184   (1)      1           1              1                   1
42586   (3)      3           3              3                   3
48575   (4)      4           4              1                   4
48431   (1)      1           1              0                   1
48555   (1)      0           1              0                   1
44208   (2)      2           2              2                   2
12880   (27)     18          6              7                   6
48944   (13)     12          0              0                   0
57357   (1)      1           1              0                   1
59442   (1)      1           1              1                   1
48289   (1)      1           1              1                   1
25124   (2)      1           1              1                   1
43754   (2)      2           2              1                   2
47796   (1)      1           1              1                   1
41900   (1)      1           1              1                   1
12660   (1)      1           0              0                   0
44375   (1)      0           0              1                   0
8571    (1)      1           0              0                   0
-----------------------------------------------------------------------------
\end{lstlisting}
        \par %
        \caption{Destination Accessibility For Accessible Networks}
        \label{lst:reachability_proxy}
    \end{figure}


{
\renewcommand{\arraystretch}{1.3}%
\begin{figure}
    {\footnotesize
        \begin{tabular}{|p{10cm}|l|}
            \hline
                Network & Addresses \\
            \hline
ASK-AS Andishe Sabz Khazar Autonomous System (39308) &    7\\
NGSAS Neda Gostar Saba Data Transfer Company Private Joint (39501) &   4\\
TIC-AS Telecommunication Infrastructure Company (48159) &      9\\
IR-PARSUN Parsun Network Solutions, IR  (31732) &       1\\
IR-AVABARID-AS Rasaneh Avabarid Private Joint Stock Company  (51431) &  1\\
AZADNET Azadnet Autonomous System  (24631) &    1\\
TEBYAN Tebyan-e-Noor Cultural-Artistic Institute  (48434) &     3\\
PAYAMAVARAN-KAVIR Shabakeh Gostar Payamavaran Kavir Company (Private Joint Stock) (57454) &    1\\
PARSONLINE PARSONLINE Autonomous System (16322) &      1\\
SINET-AS Soroush Rasanheh Company Ltd (21341) &        6\\
FARAHOOSH Farahoosh Dena (44208) &     1\\
DCI-AS Information Technology Company (ITC) (12880) &  403\\
ASKHALIJFARSONLINE Khalij Ettela Resan Jonoub LTD (48944) &    1\\
NEDA-AS neda rayaneh (30902) & 1\\
IR-PISHGAMAN-ICP Pishgaman Kavir Yazd (34918) &        2\\
HAMARA-AS Hamara System Tabriz Engineering Company (47262) &   3\\
ASIATECH-AS AsiaTech Inc. (43754) &    3\\
AFRANET AFRANET Co. Tehran, Iran (25184) &     1\\
OFOGHNET-AS Mortabet Rayaneh Ofogh (29020) &   1\\
FANAVA-AS Fanava Group (41881) &       8\\
            \hline
        \end{tabular}
        \caption{Identifiable Networks and Sites in Private IP Space Routes}
        \label{table:public_traceroute_addresses}
    }
\end{figure}
}

  \begin{figure}[h]
        \footnotesize
\begin{lstlisting}[basicstyle=\scriptsize\ttfamily,captionpos=b]
$ dig realm.blizz.ir 

; <<>> DiG 9.3.6-P1-RedHat-9.3.6-20.P1.el5_8.2 <<>> realm.blizz.ir
;; global options:  printcmd
;; Got answer:
;; ->>HEADER<<- opcode: QUERY, status: REFUSED, id: 4166
;; flags: qr rd ra; QUERY: 1, ANSWER: 0, AUTHORITY: 0, ADDITIONAL: 0

;; QUESTION SECTION:
;realm.blizz.ir.                        IN      A

;; Query time: 213 msec
;; SERVER: 8.8.8.8#53(8.8.8.8)
;; WHEN: Tue Sep 18 06:55:27 2012
;; MSG SIZE  rcvd: 32

---

$ dig realm.blizz.ir @a.irnic.ir

; <<>> DiG 9.3.6-P1-RedHat-9.3.6-20.P1.el5_8.2 <<>> realm.blizz.ir @a.irnic.ir
;; global options:  printcmd
;; Got answer:
;; ->>HEADER<<- opcode: QUERY, status: NOERROR, id: 8249
;; flags: qr rd ra; QUERY: 1, ANSWER: 1, AUTHORITY: 2, ADDITIONAL: 0

;; QUESTION SECTION:
;realm.blizz.ir.                        IN      A

;; ANSWER SECTION:
realm.blizz.ir.         3600    IN      A       10.175.27.120

;; AUTHORITY SECTION:
blizz.ir.               14400   IN      NS      ns1.blizz.ir.
blizz.ir.               14400   IN      NS      ns2.blizz.ir.

;; Query time: 32 msec
;; SERVER: 194.225.70.89#53(194.225.70.89)
;; WHEN: Tue Sep 18 06:55:16 2012
;; MSG SIZE  rcvd: 84


\end{lstlisting}
        \par %
        \caption{Failure to Propagate DNS Records}
        \label{lst:dns_prop_failure}
    \end{figure}

\newpage

\begin{figure}
    \centering
    \includegraphics[width=1\textwidth]{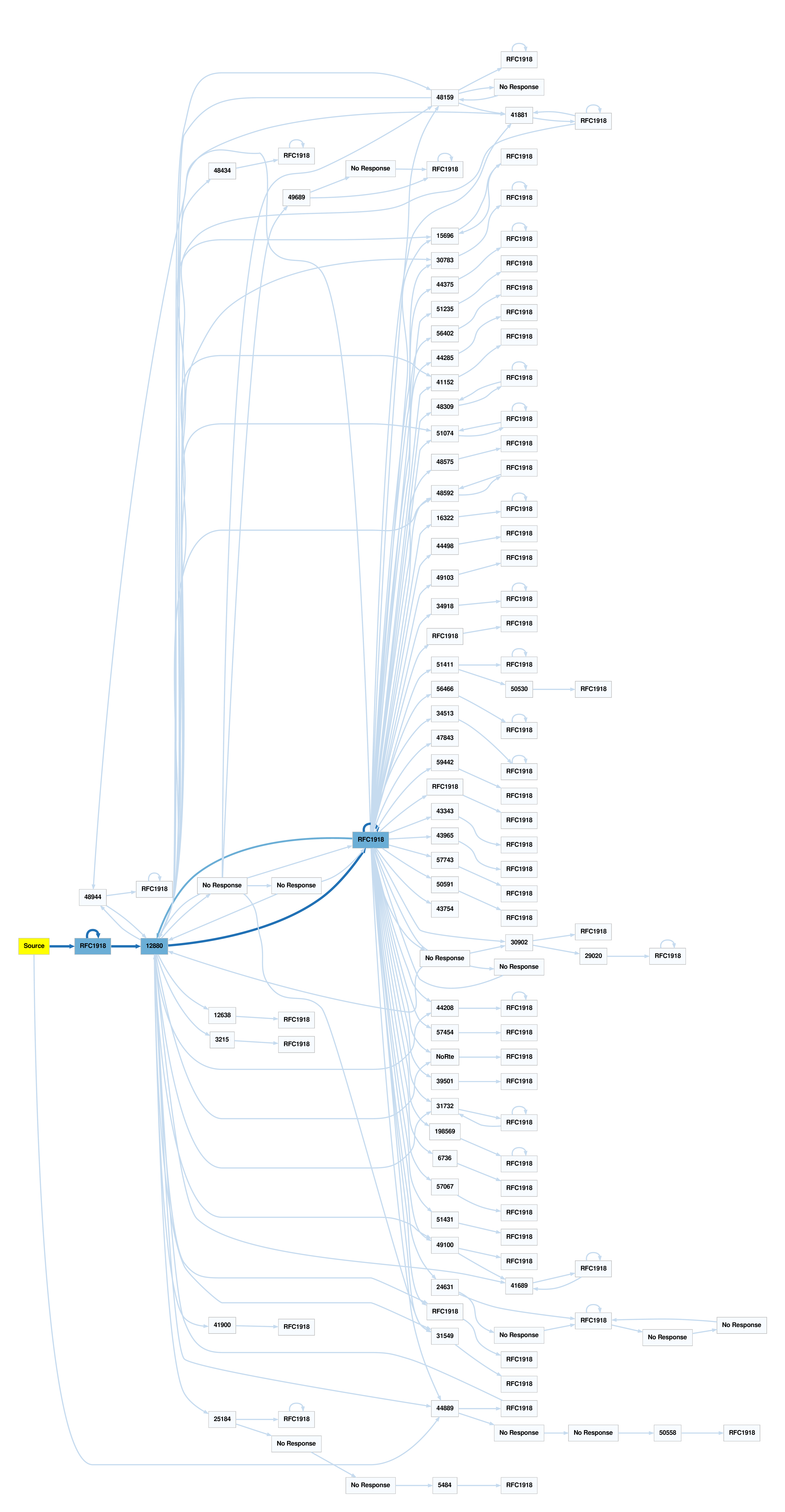}
    \caption{Traceroute Pathes From Host 1}
    \label{fig:10dotroutes_khayyam}
\end{figure}

\newpage

\begin{figure}
    \centering
    \includegraphics[width=1\textwidth]{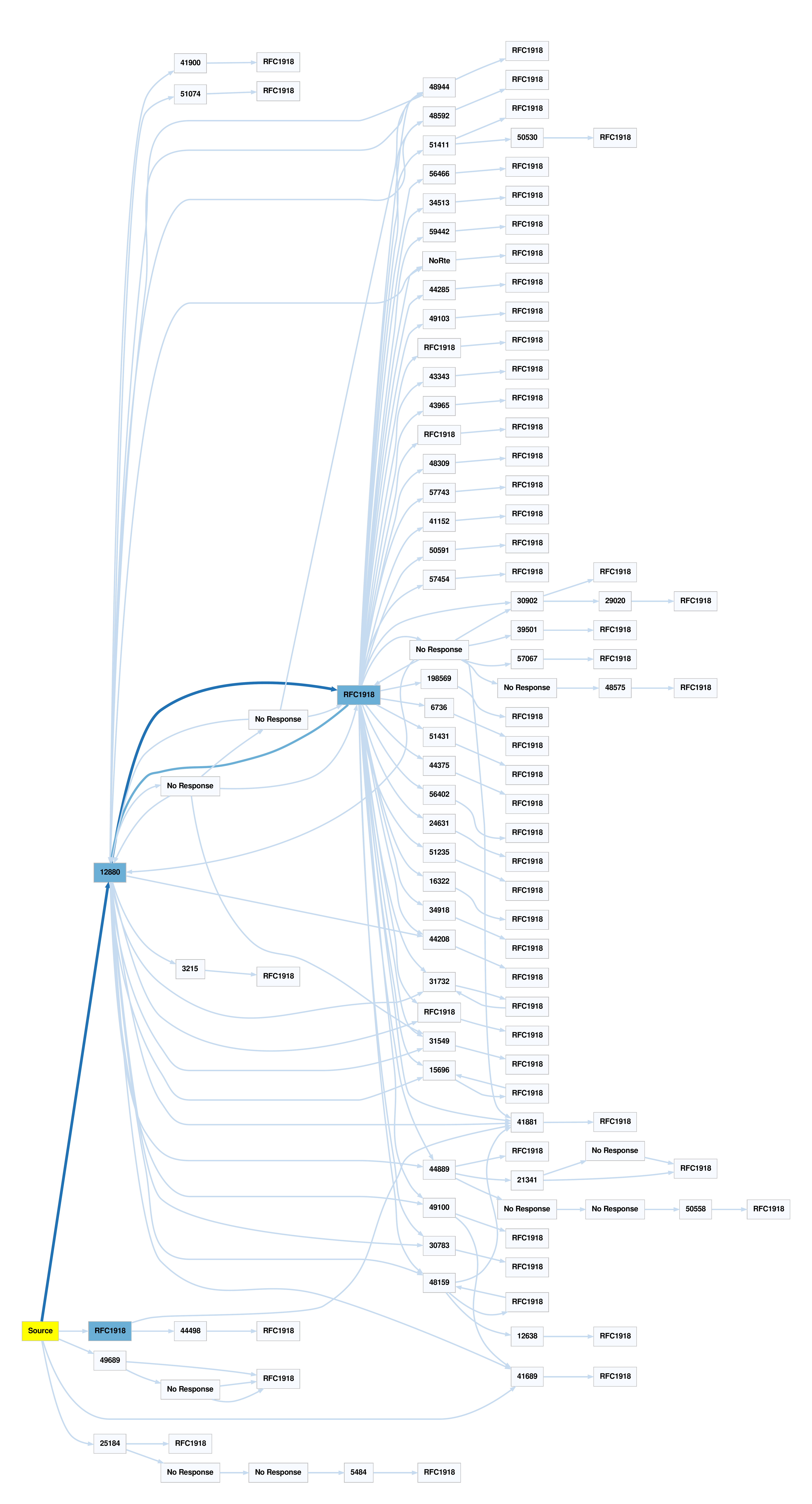}
    \caption{Traceroute Pathes From Host 2}
    \label{fig:10dotroutes_ferdowsi}
\end{figure}

\end{document}